\documentclass[sigconf]{acmart}

\usepackage{booktabs} 
\usepackage{multirow}
\usepackage{sistyle}
\SIthousandsep{,}
\usepackage{bbm}
\usepackage[british]{babel}
\usepackage{pgfplots}
\usetikzlibrary{patterns}
\usepackage{filecontents}
\pgfplotsset{compat=1.3}
\setcopyright{rightsretained}

\makeatletter
\g@addto@macro\normalsize{%
  \abovedisplayskip 0pt plus1pt 
  \belowdisplayskip
  \abovedisplayskip
  \abovedisplayshortskip  0pt plus1pt%
  \belowdisplayshortskip  0pt plus1pt
}
\makeatother

\acmDOI{10.475/123_4}

\acmISBN{123-4567-24-567/08/06}

\acmConference[WOODSTOCK'97]{ACM Woodstock conference}{July 1997}{El
  Paso, Texas USA}
\acmYear{1997}
\copyrightyear{2016}

\acmPrice{15.00}

\definecolor{green1}{RGB}{204,203,102}
\definecolor{blue1}{RGB}{102,204,182}
\definecolor{red1}{RGB}{102,163,204}
\definecolor{purple1}{RGB}{102,33,102}
\definecolor{green2}{RGB}{255,153,102}
\definecolor{blue2}{RGB}{51,204,204}
\definecolor{red2}{RGB}{143,153,204}

\definecolor{tuatara}{RGB}{67, 67, 67}
\definecolor{aluminum}{RGB}{153,153,153}
\definecolor{silver}{RGB}{191,191,191}
\definecolor{platinum}{RGB}{228,227,228}
\definecolor{mercury}{RGB}{240,240,240}
\definecolor{gallery}{RGB}{250,250,250}
\definecolor{free_speech_aquamarine}{RGB}{0, 156, 114}
\definecolor{sun_shade}{RGB}{255, 144, 68}
\definecolor{fern}{RGB}{101,197,117}
\definecolor{french_blue}{RGB}{0, 112, 182}
\definecolor{sushi}{RGB}{117, 168, 47}
\definecolor{shakespeare}{RGB}{35, 184, 223}
\definecolor{egg_shell}{RGB}{238, 234, 215}
\definecolor{carnation}{RGB}{245, 80, 86}
\definecolor{flamingo}{RGB}{237, 88, 85}
\definecolor{jet_stream}{RGB}{188, 214, 210}
\definecolor{jelly_bean}{RGB}{45, 126, 150}
\definecolor{tree_poppy}{RGB}{246, 154, 27}

\begin{document}
\title{MatchZoo: A Toolkit for Deep Text Matching}

\author{Yixing Fan, Liang Pang, JianPeng Hou, Jiafeng Guo, Yanyan Lan, Xueqi Cheng}
\affiliation{%
  \institution{CAS Key Lab of Network Data Science and Technology, Institute of Computing Technology, \\ Chinese Academy of Sciences}
  \city{Beijing, China}
}
\email{{fanyixing, pangliang, houjianpeng}@software.ict.ac.cn, {guojiafeng, lanyanyan, cxq}@ict.ac.cn}

\begin{abstract}
In recent years, deep neural models have been widely adopted for text matching tasks, such as question answering and information retrieval, showing improved performance as compared with previous methods. In this paper, we introduce the MatchZoo toolkit that aims to facilitate the designing, comparing and sharing of deep text matching models. Specifically, the toolkit provides a unified data preparation module for different text matching problems, a flexible layer-based model construction process, and a variety of training objectives and evaluation metrics. In addition, the toolkit has implemented two schools of representative deep text matching models, namely representation-focused models and interaction-focused models. Finally, users can easily modify existing models, create and share their own models for text matching in MatchZoo.



\end{abstract}

%
%



\maketitle

\section{Introduction}
Many natural language processing (NLP) tasks can be formulated as a matching problem between two texts, such as paraphrase identification, question answering, and information retrieval (IR). In recent years, a number of deep neural models have been designed and applied to such kind of tasks, e.g., DSSM~\cite{DSSM}, MatchPyramid~\cite{pang2016text} and DRMM~\cite{guo2016deep}, leading to state-of-the-art performances as compared with previous methods. However, these deep text matching models are often designed with special layers, implemented in different environments, and targeted to different matching tasks, making it difficult for reproduction and comparison.


In this paper, we introduce the MatchZoo toolkit that aims to facilitate the designing, comparing and sharing of deep text matching models. Firstly, the toolkit provides a data preparation module to convert dataset of different text matching tasks into a unified format so that they can be used as the input of different deep matching models. Secondly, the toolkit allows users to easily construct a deep matching model layer by layer based on the Keras \cite{chollet2015keras} libarary. We have extended the Keras libarary to include layer interfaces that are specifically designed for text matching problems. Moreover, the toolkit has implemented two schools of representative deep text matching models, namely representation-focused models and interaction-focused models, making it an off-the-shelf model repository. Users can directly apply these models or modify them by simple configuration. Finally, the toolkit provides a variety of training objectives for optimization, as well as a set of evaluation metrics for comparison between different deep matching models. The MatchZoo toolkit is licensed under the permissive Apache license 2.0\footnote{The toolkit is licensed under the permissive Apache open-source license and can be found at https://github.com/faneshion/MatchZoo}.

\section{The MatchZoo Toolkit}
\begin{figure}[!tbp]
\centering
\includegraphics[scale=0.4]{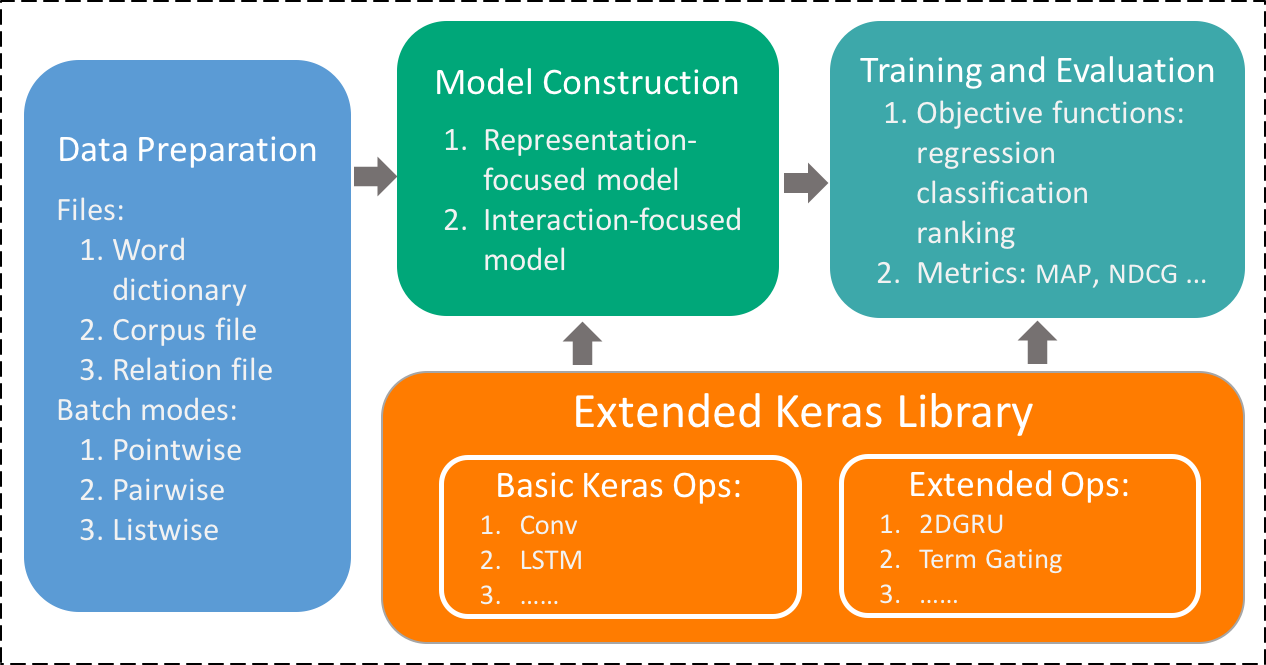}
\caption{The Architecture of the MatchZoo toolkit.}
\label{fig:matchzoo}
\end{figure}

The architecture of the MatchZoo toolkit is depicted in Figure \ref{fig:matchzoo}. There are three major modules in the toolkit, namely data preparation, model construction, training and evaluation, respectively. These three modules are actually organized as a pipeline of data flow. The toolkit is built upon the Keras library,  so that it can run over TensorFlow \cite{tensorflow2015-whitepaper}, CNTK\cite{CNTK} and Theano\footnote{http://deeplearning.net/software/theano/}, and work seamlessly with CPUs and GPUs. We will introduce each of the three modules in more details in the following.


\subsection{Data Preparation}
The data preparation module aims to convert dataset of different text matching tasks into a unified format as the input of deep matching models. Users provide datasets which contain pairs of texts along with their labels, and the module produces the following files.

\begin{itemize}
\item \textbf{Word Dictionary} records the mapping from each word to a unique identifier called \emph{wid}. Words that are too frequent (e.g. stopwords), too rare or noisy (e.g. fax numbers) can be filtered out by predefined rules.
\item \textbf{Corpus File} records the mapping from each text to a unique identifier called \emph{tid}, along with a sequence of word identifiers contained in that text. Note here each text is truncated or padded to a fixed length customized by users.
\item \textbf{Relation File} is used to store the relationship between two texts, each line containing a pair of \emph{tid}s and the corresponding label.
\end{itemize}

Beyond these basic files, users are also allowed to provide additional feature files, e.g., word embedding files or human-crafted feature files, as model input.
After converting the raw dataset to the above unified format, the module provides three types of data batch modes, i.e., generating a batch of data in pointwise, pairwise or listwise manner.

\subsection{Model Construction}
In the model construction module, we employ Keras libarary to help users build the deep matching model layer by layer conveniently. The Keras libarary provides a set of common layers widely used in neural models, such as convolutional layer, pooling layer, dense layer and so on. To further facilitate the construction of deep text matching models, we extend the Keras libarary to provide some layer interfaces specifically designed for text matching. We list a few examples here.
\begin{itemize}
\item \textbf{Matching\_Matrix} layer builds a word-by-word matching matrix based on dot product, cosine similarity or indicator function~\cite{pang2016text}.
\item \textbf{Term\_Gating} layer builds a term gating layer based on word embeddings from texts~\cite{guo2016deep}.
\item \textbf{2D-GRU} layer builds a two dimensional recurrent neural network using the Gated Recurrent Unit (GRU)~\cite{wan2016match} .
\end{itemize}
Moreover, the toolkit has implemented two schools of representative deep text matching models, namely representation-focused models and interaction-focused models~\cite{guo2016deep}. \begin{itemize}
\item \textbf{Representation-based models} include ARC-I~\cite{ARC-II}, DSSM~\cite{DSSM}, CDSSM~\cite{CDSSM}, MV-LSTM~\cite{MV-LSTM}, and CNTN~\cite{CNTN};
\item \textbf{Interaction-based models} include DRMM~\cite{guo2016deep}, ARC-II~\cite{ARC-II}, MatchPyramid\cite{pang2016text}, and Match-SRNN~\cite{wan2016match};
\end{itemize}
Users can apply these models out-of-the-box or modify them by simple configuration. Moreover, as a shared model repository, users can contribute their own matching models into this toolkit easily.


\subsection{Training and Evaluation}
For learning the deep matching models, the toolkit provides a variety of objective functions for regression, classification and ranking. For example, the ranking-related objective functions include several well-known pointwise, pairwise and listwise losses. It is flexible for users to pick up different objective functions in the training phase for optimization. Once a model has been trained, the toolkit could be used to produce a matching score, predict a matching label, or rank target texts (e.g., a document) against an input text.

For evaluation, the toolkit provides several widely adopted evaluation metrics, such as Precision, MAP, NDCG and so on. For IR tasks, the toolkit can also output the ranking results as a TREC-compatible file that can be used as the input to the trec\_eval\footnote{https://github.com/usnistgov/trec\_eval} evaluation utitlity.


\section{Conclusions}
In this paper, we introduce the MatchZoo toolkit which helps the design, comparison and share of deep text matching models. The toolkit contains three major modules, namely data preparation, model construction, training and evaluation. Meanwhile, the toolkit provides implementations of state-of-the-art deep matching models on NLP and IR related tasks. Users can easily modify these off-the-shelf models or implement their own models by using the layer interfaces provided by the toolkit.

For the future work, we would like to continuously extend the model repository and layer interfaces in the MatchZoo toolkit for better usage. It would also be interesting to accommodate other types of data beyond text for new matching scenarios.

\bibliographystyle{ACM-Reference-Format}

\end{document}